\documentclass[siam10.clo]{siamltex}

\begin{document}
\title{Computation of expansions for the maximum likelihood estimator and its distribution function}

\author{Shanti Venetiaan\thanks{Faculty of Technology, Anton de Kom Universiteit van Suriname, Paramaribo, Suriname} }

\maketitle

\begin{abstract}
In this paper, insight is given in the techniques used to compute asymptotic expansions. In a broad fashion the technique is described. Most of the results apply to the paper " An expansion for the maximum likelihood estimator and its distribution function", which will be submitted.  
\end{abstract}



\pagestyle{myheadings} \thispagestyle{plain} \markboth{ SHANTI VENETIAAN}{{\small CALCULATION OF  EXPANSIONS FOR THE MLE AND ITS DISTRIBUTION FUNCTION}}

\section{Expansions}

In the paper " An expansion for the maximum likelihood estimator and its distribution function", which will be submitted, many expansions are being calculated. Because the technique itself is not so difficult and the outcomes take a lot of space to present, many steps of the computation of the expansions will be omitted in that paper. However if one wants to check the computations, it may be useful to have some insight in how the expansions were obtained. None of the applied techniques are claimed by the author. They are just written down. We will give the full outcome of all the steps needed to construct the needed expansions. We will not give any proofs, just the method of obtaining the expansions. This means that certain rest terms will be omitted. Also note that the results are only valid for the maximum lielihood estimator of location.

\section{An expansion for the maximum likelihood estimator}
We define the maximum likelihood estimator $\hat \theta_n$ by 
\begin{equation}
L_n(\hat \theta_n)= \inf_{\theta \in  R}L_n(\theta),
\end{equation}
where $L_n(\theta) = n^{-1}\sum_{i=1}^n \rho(X_i - \theta)$, where $\rho(\cdot)= - \log(\cdot)$.
It may be proved that (cf. Chibisov (1973)) $\hat \theta_n$ is the solution of the equation
\begin{equation} L_n'(\theta) = 0,
\end{equation}
with probability $1+ o(n^{-3/2})$.
We expand $L_n'(\theta)$ with a Taylor expansion and get
\begin{eqnarray} L_n'(\theta) &=& \frac1{n}\sum_{i=1}^n \rho'(X_i-\theta)=\frac1{n}\sum_{i=1}^n \rho'(X_i) -\frac{\theta}{n}\sum_{i=1}^n \rho''(X_i)+\frac{\theta^2}{2n}\sum_{i=1}^n \rho^{(3)}(X_i)\label{eq1}\\
&\quad -&\frac{\theta^3}{6n}\sum_{i=1}^n
\rho^{(4)}(X_i)+\frac{\theta^4}{24n}\sum_{i=1}^n
\rho^{(5)}(X_i-\theta'),\nonumber\end{eqnarray} where
$|\theta'|\le |\theta|$.

We introduce the notation 
\begin{equation}\xi_j = \frac1{\sqrt{n}}\sum_{i=1}^n (\rho^{(j)}(X_i)-a_j),\quad a_j = E_0\rho^{(j)}(X)\quad\mbox{for}\quad j=1,...,5.\label{eq2}\end{equation}
Note that the $\xi_{j}'s$ are normalized sums and that $a_1 = 0$.
(\ref{eq1}) becomes 
\begin{eqnarray}
L_n'(\theta)  &=& \frac{\xi_{1n}}{\sqrt
n}-\theta(\frac{\xi_{2n}}{\sqrt n}+a_2)
+\frac{\theta^2}{2}(\frac{\xi_{3n}}{\sqrt n}+a_3)\nonumber\\
&\quad -& \frac{\theta^3}{6}(\frac{\xi_{4n}}{\sqrt
n}+a_4)+\frac{\theta^4}{24}(\frac{\xi_{5n}}{\sqrt n}+a_5)
+\cdots,\label{eq3}
\end{eqnarray}
To find the expansion for $\hat \theta_n$ we put
\begin{equation}
\hat \theta_n = B_1/\sqrt n + B_2/n + B_3/n^{3/2} + B_4/n^2.
\end{equation}
Substituting this into (\ref{eq3}) leads to
\begin{eqnarray}
&&(-a_2B_1+\xi_1)/n^{1/2}+(-\xi_2B_1-a_2B_2+{\frac12}a_3B_1^2)/n\nonumber\\
&+&(-\xi_2B_2-{\frac16}a_4B_1^3+{\frac12}\xi_3B_1^2+a_3B_1B_2-a_2B_3)/n^{3/2}\label{eq4}\\
&+&(-{\frac12}a_4B_1^2B_2+{\frac12}a_3B_2^2+{\frac{1}{24}}
a_5B_1^4-\xi_2B_3+\xi_3B_1B_2-{\frac16}\xi_4B_1^3+a_3B_1B_3-a_2B_4)/n^2\nonumber
\end{eqnarray}
Now we take the first term of (\ref{eq4}) and put it equal to 0. We get
\begin{equation}
-a_2B_1+\xi_1= 0 \Rightarrow B_1 = \xi_1/a_2.
\end{equation}
Substituting the obtained $B_1$ in (\ref{eq4}) gives
\begin{eqnarray}
&&(-\xi_2\xi_1/a_2+{\frac12}a_3\xi_1^2/a_2^2-a_2B_2)/n\label{eq5}\\
&+&(-a_2B_3+a_3\xi_1B_2/a_2-\xi_2B_2+{\frac12}\xi_3\xi_1^2/a_2^2-{\frac16}a_4\xi_1^3/a_2^3)/n^{3/2}\nonumber\\
&+&(a_3\xi_1B_3/a_2+\frac{1}{24}a_5\xi_1^4/a_2^4-{\frac12}a_4\xi_1^2B_2/a_2^2-\nonumber\\
&&\frac16 \xi_4\xi_1^3/a_2^3+\frac12 a_3B_2^2+\xi_3\xi_1B_2/a_2-\xi_2B_3-a_2B_4)/n^2\nonumber
\end{eqnarray}
Note that the $1/\sqrt n$ term has vanished. We take the first term of (\ref{eq5}) and put it equal to 0. This results in
\begin{equation}
B_2 = -\xi_1\xi_2/a_2^2+\frac12 \xi_1^2a_3/a_2^3.
\end{equation}
We substitute this $B_2$ in (\ref{eq5}) and get
\begin{eqnarray}&&(-a_2B_3-\frac16 a_4\xi_1^3/a_2^3+\frac12\xi_3\xi_1^2/a_2^2-\frac32\xi_2\xi_1^2a_3/a_2^3+\xi_2^2\xi_1/a_2^2+\frac12\xi_1^3a_3^2/a_2^4)/n^{3/2}\label{eq6}\\
&+&(-\frac16\xi_4\xi_1^3/a_2^3+a_3\xi_1B_3/a_2+\frac18\xi_1^4a_3^3/a_2^6+\frac12\xi_3\xi_1^3a_3/a_2^4\nonumber\\
&\quad-&\frac12\xi_1^3\xi_2a_3^2/a_2^5-\xi_2B_3-\xi_3\xi_1^2\xi_2/a_2^3\nonumber\\
&&+\frac{1}{24}a_5\xi_1^4/a_2^4
+\frac12 a_3\xi_1^2\xi_2^2/a_2^4+\frac12 a_4\xi_1^3\xi_2/a_2^4-a_2B_4-\frac14 a_4\xi_1^4 a_3/a_2^5)/n^2\nonumber
\end{eqnarray}
Again we put the first term of the result to 0 and get
\begin{equation}
B_3 = \xi_1\xi_2^2/a_2^3-\frac16\xi_1^3a_4/a_2^4+\frac12\xi_1^2\xi_3/a_2^3-\frac32\xi_1^2\xi_2a_3/a_2^4+\frac12\xi_1^3a_3^2/a_2^5.
\end{equation}
We substitute $B_3$ in (\ref{eq6}) to obtain
\begin{eqnarray}&&-a_2B_4-\frac52\xi_1^3\xi_2a_3^2/a_2^5+\frac58\xi_1^4a_3^3/a_2^6-\xi_2^3\xi_1/a_2^3+\xi_3\xi_1^3a_3/a_2^4
-\frac32\xi_3\xi_1^2\xi_2/a_2^3\\
&&+\frac23a_4\xi_1^3\xi_2/a_2^4+\frac{1}{24}a_5\xi_1^4/a_2^4-\frac16\xi_4\xi_1^3/a_2^3+3a_3\xi_1^2\xi_2^2/a_2^4-\frac{5}{12}a_4\xi_1^4a_3/a_2^5.\nonumber
\end{eqnarray}
This is put equal to 0 and at last we obtain
\begin{eqnarray}
B_4&=& \xi_1^3\xi_3a_3/a_2^5-\frac52\xi_1^3\xi_2a_3^2/a_2^6+\frac58\xi_1^4a_3^3/a_2^7-\xi_1\xi_2^3/a_2^4
-\frac16\xi_1^3\xi_4/a_2^4\\
&&-\frac32\xi_1^2\xi_3\xi_2/a_2^4+\frac23\xi_1^3a_4\xi_2/a_2^5+\frac{1}{24}\xi_1^4a_5/a_2^5+3\xi_1^2a_3\xi_2^2/a_2^5-\frac{5}{12}\xi_1^4a_4a_3/a_2^6.\nonumber
\end{eqnarray}
Eventually 
\begin{eqnarray}\sqrt n(\hat \theta_n) &=& \frac{\xi_1}{a_2}
+\frac1{\sqrt{n}}(\frac{-\xi_1\xi_2}{a_2^2}+\frac{a_3\xi_1^2}{2a_2^3})+ \frac1{n}(\frac{\xi_1\xi_2^2}{a_2^3} - \frac{3a_3\xi_1^2\xi_2}{2a_2^4}+\frac{\xi_1^2\xi_3}{2a_2^3}+\frac{a_3^2\xi_1^3}{2a_2^5}-\frac{a_4\xi_1^3}{6a_2^4})\nonumber\\
&\quad + &\frac1{n^{3/2}}\biggl(\frac{3\xi_1^2a_3\xi_2^2}{a_2^5}+\frac{5\xi_1^4a_3^3}{8a_2^7}-\frac{5\xi_1^4a_4a_3}{12a_2^6}-\frac{3\xi_1^2\xi_3\xi_2}{2a_2^4}-\frac{5\xi_1^3a_3^2\xi_2}{2a_2^6}+\frac{\xi_1^4a_5}{24a_2^5}\label{eq7}\\
&\qquad + &\frac{\xi_1^3\xi_3a_3}{a_2^5}+\frac{2\xi_1^3a_4\xi_2}{3a_2^5}-\frac{\xi_1^3\xi_4}{6a_2^4}-\frac{\xi_1\xi_2^3}{a_2^4}\biggr)+\cdots,\nonumber
\end{eqnarray} 
	
\section{Expansion for the distribution function of the maximum likelihood estimator}	
The estimator in (\ref{eq7}) fits the model of Hall (1992), Section 2.3. This means that the cumulants of $\sqrt n\hat\theta_n$ will determine the expansion for the distribution function.
First we note that the cumulants will be of the form (cf. Hall(1992), Section 2.3)
\begin{eqnarray}
\kappa_1 &=& 0 + k_{12}/{\sqrt n} + k_{13}/n^{3/2}+\cdots\nonumber\\
\kappa_2 &=& k_{21}+ k_{22}/n +\cdots \nonumber\\
\kappa_3 &=& k_{31}/{\sqrt n} + k_{32}/n^{3/2}+\cdots\label{eq8}\\
\kappa_4 &=& k_{41}/n + \cdots\nonumber\\
\kappa_5 &=& k_{51}/n^{3/2}+\cdots.\nonumber
\end{eqnarray}
Furthermore (cf. Hall(1992), Section 2.2)
\begin{eqnarray}
\kappa_1 &=& ES_n\nonumber\\
\kappa_2&=& ES_n^2 - (ES_n)^2\nonumber\\
\kappa_3 &=& E(S_n-ES_n)^3= ES_n^3 - 3ES_n^2ES_n + 2(ES_n)^3\label{eq9}\\
\kappa_4&=& E(S_n - ES_n)^4 - 3\kappa_2^2\nonumber\\
\kappa_5 &=& E(S_n - ES_n)^5 - 10\kappa_2\kappa_3.\nonumber
\end{eqnarray}

\subsection{Computation of the kappa's}
We will now compute the kappa's.
For the computation of the kappa's we need to calculate the expectation of the normalized sums, the so called $\xi_j$'s. Terms that become too small will be omitted. 
First we will introduce the notation
\begin{equation}\psi_i(\cdot) = \frac{f^{(i)}}{f}(\cdot) \end{equation}

\begin{eqnarray}\eta_2 &=& E(\psi_2^2(X_i)),\quad \eta_3 = E(\psi_1^3(X_i)),\quad \eta_4 = 
E(\psi_1^4(X_i)),\\
\eta_5 &=& E(\psi_1^5(X_i)),\quad \eta_6 = E(\psi_2\psi_3(X_i))\nonumber\end{eqnarray}

\ the above results in
\begin{eqnarray}a_1 &=&0,\quad a_2 = E(\psi_1^2(X_i)),\mbox{without loss of generality we put}\quad  a_2 =1,\\ a_3 &=& -\frac12\eta_3,\quad a_4 = \frac23 \eta_4- \eta_2 ,\quad a_5 =5\eta_6 -\frac32\eta_5 .\nonumber
\end{eqnarray}
and that
\[E(\psi_1\psi_2)=\frac12\eta_3,\quad E(\psi_1\psi_3)=-\eta_2+\frac23\eta_4,\quad 
E(\psi_1\psi_4)=-5\eta_6+\frac32\eta_5
\]
\[E(\psi_1^2\psi_2)=\frac23\eta_4,\quad E(\psi_1\psi_2^2)=2\eta_6,\quad E(\psi_1^3\psi_2)=\frac34\eta_5,\quad E(\psi_1^2\psi_3)=-4\eta_6+\frac32\eta_5 \]

Furthermore
\[ w_j = -( \rho^{(j)}(X_i) - a_j), \mbox{for}\quad j = 1,...,5.\]
Consequently,
\begin{eqnarray}
w_1 &=& \psi_1,\\
w_2 &=& \psi_2- \psi_1^2+1,\nonumber\\
w_3&=&\psi_3-3\psi_1\psi_2+2\psi_1^3+a_3,\nonumber\\
w_4&=&\psi_4-4\psi_1\psi_3+12\psi_1^2\psi_2-3\psi_2^2-6\psi_1^4+a_4\nonumber\\
w_5 &=& \psi_5-5\psi_1\psi_4+20\psi_1^2\psi_3-10\psi_2\psi_3+30\psi_1\psi_2^2-60\psi_1^3\psi_2+24\psi_1^5+a_5\nonumber
\end{eqnarray}
Note that $Ew_j =0$ for $j=1,\cdots 5$ and that $E(w_1^2) =1$.

\[E(w_1w_2) = -\frac12 \eta_3,\quad E(w_1w_3)=\frac23\eta_4-\eta_2,\quad E(w_1w_4)=5\eta_6-\frac32\eta_5
\]

\[E(w_1^2w2)=-\frac13\eta_4+1,\quad E(w_1^2w_3)=-4\eta_6+\frac54\eta_5-\frac12\eta_3 \]

\[E(w_1^3w_2)=-\frac14\eta_5+\eta_3,\quad E(w_1w_2^2)=2\eta_6-\frac12\eta_5-\eta_3\]

\[E(w_2^2) = \eta_2-\frac13\eta_4-1,\quad E(w_2w_3)= -\eta_6+\frac14\eta_5+\frac12\eta_3\]

We will give an example here to illustrate how the expectations of the terms of $\sqrt n\hat\theta_n$ are obtained.

\begin{eqnarray}
E\xi_1^2&=& E(\frac1{\sqrt n}\sum_i w_{1i})^2\nonumber \\
&=& E\{\frac1n (\sum_i w_{1i}^2+2\sum\sum_{i<j} w_{1i}w_{1j}) \}\\
&=& \frac1n(n Ew_1^2+\frac{2n(n-1)}{2}Ew_1Ew_1) = Ew_1^2+ 0 = Ew_1^2=1.\nonumber
\end{eqnarray}

\begin{eqnarray}
E\xi_1^8 &=& E(\frac1{\sqrt n}\sum_i w_{1i})^8\\
&=& \frac1{n^4} E\Bigl[\sum_i w_{1i}^8 + 8\sum\sum_{i\neq j} w_{1i}^7w_{1j}+{8\choose 2}
\sum\sum_{i\neq j}w_{1i}^6w_{1j}^2\nonumber\\
&\quad +&{8\choose 3}\sum\sum_{i\neq j} w_{1i}^5w_{1j}^3+{8\choose 4}\sum\sum_{i < j} w_{1i}^4
w_{1j}^4\nonumber\\
&\quad +& {8\choose{4,2,2}}\sum\sum\sum_{i\neq j<k\neq i}w_{1i}^4w_{1j}^2w_{1k}^2 +{8\choose{2,3,3}}\sum\sum\sum_{i\neq j<k\neq i}w_{1i}^2w_{ij}^3w_{1k}^3\nonumber\\
&\quad +& {8 \choose{2,2,2,2}}\sum\sum\sum\sum_{i<j<k<l}w_{1i}^2w_{1j}^2w_{1k}^2w_{1l}^2\Bigr]\nonumber\\
&=& \frac1{n^4}\Bigl[{8 \choose{2,2,2,2}}\frac{n(n-1)(n-2)(n-3)}{4!}(Ew_1^2)^4\nonumber\\
&\quad+&{8 \choose{2,2,4}}\frac{n(n-1)(n-2)}{2!}Ew_1^4(Ew_1^2)^2\nonumber\\
&\quad +&{8\choose{2,3,3}}\frac{n(n-1)(n-2)}{2!}(Ew_1^3)^2Ew_1^2+...\Bigr]\nonumber\\
&=& \frac1{n^4}\Bigl[105(n^4-6n^3+11n^2-6n)(Ew_1^2)^4+ 210(n^3-3n^2+2n)Ew_1^4(Ew_1^2)^2\nonumber\\
&\quad+& 280(n^3-3n^2+2n)(Ew_1^3)^2Ew_1^2+... \Bigr] \nonumber\\
&=& 105(Ew_1^2)^4 + \frac1{n}[-630(Ew_1^2)^4 + 210Ew_1^4(Ew_1^2)^2+280(Ew_1^3)^2Ew_1^2]+...\nonumber                                                                                                                                                                                                                          \\
&=& 105 +\frac1n[-630 + 210\eta_4+280\eta_3^2]+\cdots.\nonumber
\end{eqnarray}
 
The following equations give formula's for the other expectation that have to be computed.

\begin{eqnarray}
E(\xi_1^{2k}) &=& {{2k}\choose{2,\cdots, 2}}\frac1{k!} (E(w_1^2))^k
+ \frac1{n} \Bigl[ -{{2k}\choose{2,\cdots, 2}}\frac{{{k}\choose{2}}}{k!}(Ew_1^2)^k \\
&+& {{2k}\choose{4, 2,\cdots, 2}}
\frac1{(k-2)!}Ew_1^4(Ew_1^2)^{k-2}\nonumber\\
&\quad+&{{2k}\choose{3, 3, 2, \cdots, 2}}\frac12\frac1{(k-3)!}(Ew_1^3)^2(Ew_1^2)^{k-3}  \Bigr] +\cdots\nonumber
\end{eqnarray}

\begin{eqnarray}
&&E(\xi_1^{2k+1})  = -\frac1{\sqrt n}\Bigl[{{2k+1}\choose{3, 2, \cdots, 2}}\frac{1}{(k-1)!}(Ew_1^2)^{k-1}Ew_1^3\Bigr]\\
& -&\frac1{n\sqrt n}\Bigl[-{{k}\choose{2}}{{2k+1}\choose{3, 2,\cdots, 2}}
\frac1{(k-1)!} (Ew_1^2)^{k-1}Ew_1^3\nonumber\\
&+&{{2k+1}\choose{5, 2, \cdots, 2}}\frac1{(k-2)!}(Ew_1^2)^{k-2}Ew_1^5\nonumber\\
& +& {{2k+1}\choose{4, 3, 2, \cdots, 2}}\frac1{(k-3)!}(Ew_1^2)^{k-3}Ew_1^4Ew_1^3\Bigr]+\cdots\nonumber
\end{eqnarray}

For $j=2,...4$ we have
\begin{eqnarray}
&&E(\xi_1^{2k}\xi_j)  = -\frac1{\sqrt n}\Bigl[{{2k}\choose{2,\cdots, 2}}\frac1{(k-1)!}Ew_1^2w_j(Ew_1^2)^{k-1}\\
&+&{{2k}\choose{1, 3, 2, \cdots, 2}}
\frac1{(k-2)!}Ew_1^3Ew_1w_j(Ew_1^2)^{k-2}\Bigr]\nonumber\\
& -&\frac1{n\sqrt n}\Bigl[-\frac{{{k}\choose{2}}}{(k-1)!}{{2k}\choose{2, \cdots, 2}}Ew_1^2w_j(Ew_1^2)^{k-1}\nonumber\\
&-&\frac{{{k}\choose{ 2}}}{(k-2)!}
{{2k}\choose{1, 3, 2,\cdots, 2}}Ew_1^3Ew_1w_j(Ew_1^2)^{k-2}\nonumber\\
&+& {{2k}\choose{5, 1, 2,\cdots, 2}}\frac1{(k-3)!}Ew_1^5Ew_1w_j(Ew_1^2)^{k-3}\nonumber\\
 &+& {{2k}\choose{4, 2,\cdots, 2}}
\frac1{(k-3)!}Ew_1^4Ew_1^2w_j(Ew_1^2)^{k-3}\nonumber\\
&+& {{2k}\choose{3, 3, 2,\cdots, 2}}\frac1{(k-3)!}Ew_1^3Ew_1^3w_j(Ew_1^2)^{k-3}\nonumber\\
& +& {{2k}\choose{4 ,2,\cdots, 2}}
\frac1{(k-2)!}Ew_1^4w_j(Ew_1^2)^{k-2}\Bigr]+\cdots\nonumber
\end{eqnarray}

\begin{eqnarray}
E(\xi_1^{2k+1}\xi_2) &=&  (2k+1)(2k-1)\cdots 1(E(w_1^2))^kEw_1w_2\\
&+&\frac1{n} \Bigl[ -\frac{{{k+1}\choose{2}}}{k!}{{2k+1}\choose{1, 2,\cdots, 2}}Ew_1w_2(Ew_1^2)^k\nonumber\\
&\quad+&\frac1{(k-2)!}
{{2k+1}\choose{4, 1,\cdots, 2}}Ew_1^4Ew_1w_2(Ew_1^2)^{k-2}\nonumber\\
& +&\frac12\frac1{(k-3)!}{{2k+1}\choose{3, 3, 1, 2,\cdots, 2}}(Ew_1^3)^2Ew_1w_2(Ew_1^2)^{k-3}\nonumber\\
&\quad+&\frac1{(k-1)!}
{{2k+1}\choose{3, 2,\cdots, 2}} Ew_1^3w_2(Ew_1^2)^{k-1}\nonumber\\
&& +\frac1{(k-2)!}{{2k+1}\choose{3, 2,\cdots, 2}}Ew_1^3Ew_1^2w_2(Ew_1^2)^{k-2}\Bigr]+\cdots\nonumber
\end{eqnarray}

\begin{eqnarray}
E(\xi_1^{2k}\xi_2^2) &=&  {{2k}\choose{2,\cdots, 2}}\frac1{k!}(Ew_1^2)^kEw_2^2\\
&\qquad+&{{2k}\choose{1, 1, 2,\cdots, 2}}
\frac1{(k-1)!}(Ew_1^2)^{k-1}(Ew_1w_2)^2\nonumber\\
&+&\frac1{n}\Bigl[{{2k}\choose{4,2,\cdots, 2}}\frac1{(k-2)!}Ew_1^4(Ew_1^2)^{k-2}Ew_2^2\nonumber\\
&\quad+& {{2k}\choose{3, 3, 2,\cdots, 2}}
\frac1{2\cdot (k-3)!}(Ew_1^3)^2(Ew_1^2)^{k-3}Ew_2^2\nonumber\\
&\quad+&{{2k}\choose{4, 1, 1, 2,\cdots, 2}}\frac1{(k-3)!}Ew_1^4(Ew_1w_2)^2(Ew_1^2)^{k-3}\nonumber\\
&\quad+& {{2k}\choose{1, 3, 2,\cdots, 2}}
\frac1{(k-2)!}Ew_1^3Ew_1w_2^2(Ew_1^2)^{k-2}\nonumber\\
&\quad+&{{2k}\choose{1, 3, 2,\cdots, 2}}\frac{{{2}\choose{1}}}{(k-2)!}Ew_1^3Ew_1^2w_2Ew_1w_2(Ew_1^2)^{k-2}\nonumber\\
&\quad+& {{2k}\choose{1, 3, 2,\cdots, 2}}\frac{{{2}\choose{1}}}{(k-2)!}Ew_1^3w_2Ew_1w_2(Ew_1^2)^{k-2}\nonumber\\
&\quad-&{{2k}\choose{2,\cdots, 2}}\frac{{{k+1}\choose{2}}}{k!}(Ew_1^2)^kEw_2^2\nonumber\\
&\quad+& {{2k}\choose{2,\cdots, 2}}
\frac{{{2}\choose{1}}}{2(k-2)!}(Ew_1^2w_2)^2(Ew_1^2)^{k-2}\nonumber\\
&\quad-& {{2k}\choose{1, 1, 2, \cdots, 2}}\frac{{{k+1}\choose{2}}}{(k-1)!}(Ew_1^2)^{k-1}(Ew_1w_2)^2\nonumber\\
&\quad+& {{2k}\choose{2,\cdots, 2}}\frac1{(k-1)!}(Ew_1^2w_2^2(Ew_1^2)^{k-1}\Bigr]+\cdots\nonumber
\end{eqnarray}

\begin{eqnarray}
E(\xi_1^{2k}\xi_2\xi_3)&=& {{2k}\choose{2,\cdots, 2}}\frac1{k!}(E(w_1^2)^k Ew_2w_3\\
&\quad +& \frac1{(k-1)!}{{2k}\choose{1, 1, 2\cdots 2}}Ew_1w_2Ew_1w_3+\cdots\nonumber
\end{eqnarray}

\begin{eqnarray}
E(\xi_1^{2k+1}\xi_2\xi_3)&=& -\frac1{\sqrt{n}}\Bigl[{{2k+1}\choose{3, 2,\cdots, 2}}
\frac1{(k-1)!}Ew_1^3Ew_2w_3(Ew_1^2)^{k-1}\\
&+&{{2k+1}\choose{1, 1, 3, 2, \cdots, 2}}\frac1{(k-2)!}Ew_1^3Ew_1w_2Ew_1w_3(Ew_1^2)^{k-2}\nonumber\\
&+& {{2k+1}\choose{1, 2, \cdots, 2}}\frac1{(k-1)!}Ew_1^2w_2Ew_1w_3(Ew_1^2)^{k-1}\nonumber\\
&+& {{2k+1}\choose{1, 2, \cdots, 2}}
\frac1{(k-1)!}Ew_1w_2Ew_1^2w_3(Ew_1^2)^{k-1}\nonumber\\
&+&{{2k+1}\choose{1, 2,\cdots, 2}}\frac1{k!}Ew_1w_2w_3(Ew_1^2)^k\Bigr]+\cdots.\nonumber
\end{eqnarray}

\begin{eqnarray}
E(\xi_1^{2k+1}\xi_2^2) & =& -\frac1{\sqrt n} \Bigl[{{2k+1}\choose{3, 2, \cdots, 2}}\frac1{(k-1)!}Ew_1^3Ew_2^2(Ew_1^2)^{k-1}\\
&+& {{2k+1}\choose{1, 1, 3, 2,\cdots, 2}}\frac1{(k-2)!}Ew_1^3(Ew_1w_2)^2(Ew_1^2)^{k-2}\nonumber\\
&+& {{2k+1}\choose{1, 2, \cdots, 2}}\frac1{k!}Ew_1w_2^2(Ew_1^2)^k\nonumber\\
&+&{{2k+1}\choose{1, 2, \cdots, 2}}
\frac{{{2}\choose{1}}}{(k-1)!}Ew_1^2w_2Ew_1w_2(Ew_1^2)^{k-1}\Bigr]+\cdots\nonumber
\end{eqnarray}

\begin{eqnarray}
E(\xi_1^{2k+1}\xi_2^3)& =& {{2k+1}\choose{1, 2, \cdots, 2}}\frac{{3\choose 1}}{k!}Ew_1w_2Ew_2^2(Ew_1^2)^k\\
 &\quad +& {{2k+1}\choose{1, 1, 1, 2, \cdots, 2}}{{3}\choose{1 1 1}}\frac1{3!(k-1)!}(Ew_1w_2)^3(Ew_1^2)^{k-1}+\cdots\nonumber
\end{eqnarray}

\begin{eqnarray}
E(\xi_1^{2k}\xi_2^3) & =& -\frac1{\sqrt n}\Bigl[ {{2k}\choose{1, 3, 2,\cdots, 2}}\frac{{3\choose 1}}{(k-2)!}Ew_1^3Ew_2^2Ew_1w_2(Ew_1^2)^{k-2}\\
&\qquad +& \frac1{k!}{{2k}\choose{2,\cdots, 2}}Ew_2^3(Ew_1^2)^k\nonumber \\
&\qquad +& {{2k}\choose{2,\cdots, 2}}\frac3{(k-1)!}Ew_1^2w_2Ew_2^2(Ew_1^2)^{k-1}\nonumber\\
&\qquad +& \frac{1}{(k-3)!}\cdot{{2k}\choose{1, 1, 1, 3, 2,\cdots, 2}}Ew_1^3(Ew_1w_2)^3(Ew_1^2)^{k-3}\nonumber\\
&\qquad + &{{2k}\choose{1, 1, 2,\cdots, 2}}\frac{3}{(k-1)!}(Ew_1^2)^{k-1}Ew_1w_2^2Ew_1w_2\nonumber\\
&\qquad +& {{2k}\choose{1, 1, 2,\cdots, 2}}\frac{3}{(k-2)!}(Ew_1^2)^{k-2}Ew_1^2w_2(Ew_1w_2)^2\Bigr]+\cdots\nonumber
\end{eqnarray}

\begin{eqnarray}
E(\xi_1^{2k}\xi_2^4) & =& {{2k}\choose{2,\cdots, 2}}\frac3{k!}(Ew_2^2)^2(Ew_1^2)^k\\
&\qquad +& {{2k}\choose{1, 1, 2,\cdots, 2}}\frac6{(k-1)!}(Ew_1^2)^{k-1}Ew_2^2(Ew_1w_2)^2\nonumber\\
&\qquad +& {{2k}\choose{1, 1, 1, 1, 2,\cdots, 2}}\frac1{(k-2)!}(Ew_1^2)^{k-2}(Ew_1w_2)^4+\cdots\nonumber
\end{eqnarray}

Further computations lead to

\begin{eqnarray}
ES_n &=& \frac{\eta_3}{4\sqrt n}
+\frac1{n^{3/2}}(\frac19\eta_4\eta_3+\frac1{16}\eta_5-\frac14\eta_3+\frac5{24}\eta_3\eta_2-\frac{11}{64}\eta_3^3-\frac38\eta_6)+\cdots\\
ES_n^2 &=& 1+\frac1{n}(-\frac1{16}\eta_3^2-\frac13\eta_4+\eta_2-1)+\cdots\nonumber\\
ES_n^3 &=& \frac{5\eta_3}{4\sqrt n}+\frac1{n^{3/2}}(-\frac5{12}\eta_4\eta_3+\frac{35}8\eta_3\eta_2-\frac{45}{32}\eta_3^3-\frac{15}4\eta_3-\frac{45}8\eta_6+\frac{21}{16}\eta_5)+\cdots\nonumber\\
ES_n^4 &=& 3+\frac1{n}(10\eta_2-9+\frac18\eta_3^2-\frac{11}3\eta_4)+\cdots\nonumber\\
ES_n^5 &=& \frac{35\eta_3}{4\sqrt n}+\frac1{n^{3/2}}(-\frac{175}{12}\eta_4\eta_3-\frac{525}8\eta_6-\frac{875}{64}\eta_3^3+\frac{259}{16}\eta_5+\frac{525}8\eta_3\eta_2-\frac{105}2\eta_3)\nonumber\\
&\quad+&\cdots\nonumber
\end{eqnarray}

Finally, by using (\ref{eq9}) we get the cumulants,

\begin{eqnarray}
\kappa_1 &=& \frac{\eta_3}{4\sqrt n}+\frac1{n^{3/2}}(\frac19\eta_4\eta_3+\frac1{16}\eta_5-\frac14\eta_3+\frac5{24}\eta_3\eta_2-\frac{11}{64}\eta_3^3-\frac38\eta_6)+\cdots\\
\kappa_2 &=& 1+\frac1n(-\frac18\eta_3^2-1-\frac13\eta_4+\eta_2)+\cdots\nonumber\\
\kappa_3 &=& \frac{\eta_3}{2\sqrt n}+\frac1{n^{3/2}}(-\frac94\eta_3+3\eta_3\eta_2-\frac12\eta_3\eta_4+\frac98\eta_5-\frac{13}{16}\eta_3^3-\frac92\eta_6)+\cdots\nonumber\\
\kappa_4 &=& \frac1n(-3-\frac53\eta_4+4\eta_2)+\cdots\nonumber\\
\kappa_5 &=& \frac1{n^{3/2}}(-15\eta_6-10\eta_3-5\eta_3\eta_4+15\eta_3\eta_2+4\eta_5-\frac{15}8\eta_3^3)+\cdots.\nonumber
\end{eqnarray}

Note that indeed they are of the form in (\ref{eq8}). 

\subsection{Finding the polynomials of the expansion}
We will now find the polynomials which make up the expansion of the distribution function.

We view the characteristic function as
\begin{equation}\exp \{\kappa_1 (it)+\frac12\kappa_2 (it)^2 + \frac16\kappa_3(it)^3+\frac1{24}\kappa_4(it)^4+\frac1{120}\kappa_5(it)^5+...\}
\end{equation}
By using (\ref{eq8}) we get

\begin{eqnarray}
&& \exp\{(it)(\frac{k_{12}}{\sqrt n}+\frac{k_{13}}{n\sqrt n})+\frac12(it)^2(1+\frac{k_{22}}{n})+\frac16(it)^3(\frac{k_{31}}{\sqrt n}
+\frac{k_{32}}{n\sqrt n})\nonumber\\
&+&\frac1{24}(it)^4\frac{k_{41}}{n}
+\frac1{120}(it)^5\frac{k_{51}}{n\sqrt n}  \}\nonumber\\
&=&\exp\{-\frac12 t^2\}\exp\{\frac1{\sqrt n}(k_{12}(it)+\frac16k_{31}(it)^3)+\frac1n(\frac12k_{22}(it)^2\nonumber\\
&+&\frac1{24}k_{41}(it)^4)+\frac1{n^{3/2}}(k_{13}(it)+\frac16k_{32}(it)^3+\frac1{120}k_{51}(it)^5) \}\nonumber
\end{eqnarray}

Constructing a Taylor expansion for the above gives

\begin{eqnarray}
&&\exp\{-\frac12 t^2 \}\exp\{\frac1{\sqrt n}(k_{12}(it)+\frac16k_{31}(it)^3)+\frac1n(\frac12k_{22}(it)^2\\
&\quad+&\frac1{24}k_{41}(it)^4)+\frac1{n^{3/2}}(k_{13}(it)+\frac16k_{32}(it)^3+\frac1{120}k_{51}(it)^5) \}\nonumber\\
&=& \exp\{-\frac12 t^2 \}\Bigl[1+\frac1{\sqrt n}(k_{12}(it)+\frac16k_{31}(it)^3)+\frac1n(\frac12k_{22}(it)^2\nonumber\\
&\quad+&\frac1{24}k_{41}(it)^4)+\frac1{n^{3/2}}(k_{13}(it)+\frac16k_{32}(it)^3+\frac1{120}k_{51}(it)^5)\nonumber\\
&\quad+& \frac12(\frac1n(k_{12}(it)+\frac16k_{31}(it)^3))^2)\nonumber\\
&\quad+& \frac2{n^{3/2}}(k_{12}(it)+\frac16k_{31}(it)^3))(\frac12k_{22}(it)^2+\frac1{24}k_{41}(it)^4+..)\nonumber\\
&\quad+& \frac16(\frac1{n^{3/2}}(k_{12}(it)+\frac16k_{31}(it)^3)^3) \Bigr]\nonumber\\
&=& \exp\{-\frac12 t^2 \}(1+\frac{r_1(it)}{\sqrt n}+\frac{r_2(it)}{n}+\frac{r_3(it)}{n\sqrt n}).\nonumber
\end{eqnarray}

Consequently,

\begin{eqnarray}
r_1(it) &=& ((it)k_{12}+\frac16(it)^3k_{31})\\
r_2(it)&=& (\frac12(it)^2k_{12}^2+\frac1{72}(it)^6 k_{31}^2+\frac16(it)^4k_{12}k_{31}+\frac1{24}(it)^4k_{41}+\frac12(it)^2k_{22})\\
r_3(it) &=& (\frac1{72}(it)^7k_{12}k_{31}^2+\frac1{24}(it)^5k_{12}k_{41}+\frac1{12}(it)^5k_{22}k_{31}+\frac1{144}(it)^7k_{31}k_{41}\\
&\quad+& \frac1{120}k_{51}(it)^5+\frac1{12}(it)^5k_{12}^2k_{31}+\frac12(it)^3k_{12}k_{22}\nonumber\\
&\quad+& \frac16(it)^3k_{32}+(it)k_{13}+\frac16(it)^3k_{12}^3+\frac1{1296}(it)^9k_{31}^3)\nonumber
\end{eqnarray}

The polynomials of the expansions are now( cf. Hall(1992), Section 2.2)

\begin{eqnarray}
p_1&=& -\frac1{12}\eta_3(x^2+2)\\
p_2 &=& -\frac1{288}\eta_3^2 x^5+(\frac1{72}\eta_3^2+\frac18+\frac5{72}\eta_4-\frac16\eta_2)x^3+(\frac18+\frac1{24}\eta_3^2-\frac1{24}\eta_4)x\\
p_3 &=& -\frac1{10368}\eta_3^3x^8+(\frac1{96}\eta_3+\frac{19}{10368}\eta_3^3-\frac1{72}\eta_3\eta_2+\frac5{864}\eta_4\eta_3)x^6\\
&+&(\frac{19}{1728}\eta_3^3-\frac1{30}\eta_5+\frac18\eta_6-\frac1{72}\eta_4\eta_3)x^4+(-\frac5{96}\eta_4\eta_3+\frac{35}{864}\eta_3^3+\frac1{32}\eta_3+\frac1{80}\eta_5)x^2\nonumber\\
&+& \frac{35}{432}\eta_3^3-\frac5{48}\eta_4\eta_3+\frac1{40}\eta_5+\frac1{16}\eta_3\nonumber
\end{eqnarray}

The expansion for the distribution function of the maximum likelihood estimator is now

\begin{equation}
G_n(x) = \Phi(x) + \frac1{\sqrt n}p_1(x)\phi(x) + \frac1n p_2(x)\phi(x) + \frac1{n^{3/2}}p_3(x)\phi(x)+ o(n^{-3/2}),\label{eq10}
\end{equation}
with $G_n(x) = P_0(\sqrt n\hat \theta_n\le x)$.

\subsection{Finding the Cornish-Fisher expansion for $G_n^{-1}(u)$}

We will now find a Cornish-Fisher expansion for $G_n^{-1}(\cdot)$.

Assume that $G_n^{-1}$ is of the form

\begin{equation}
G_n^{-1}(u) = z_u + \frac{A}{\sqrt n} +\frac{B}{n}+\frac{C}{n^{3/2}},
\end{equation}

where $z_u$ denotes $\Phi^{-1}(u)$. We construct Taylor expansions for the terms at the right hand side of (\ref{eq10}) and plug $G_n^{-1}(u)$ into these expansions.

The first term of (\ref{eq10})

\begin{eqnarray}
\Phi(G_n^{-1}(u)) &=& \Phi(z_u + \frac{A}{\sqrt n} +\frac{B}{n} + \frac{C}{n^{3/2}})\label{eq11}\\
&=& \Phi(z_u) + (\frac{A}{\sqrt n} +\frac{B}{n} + \frac{C}{n^{3/2}})\phi(z_u)+\frac12(\frac{A}{\sqrt n} +\frac{B}{n} )^2\cdot -z_u\phi(z_u)\nonumber\\
&\quad\quad+& \frac16(\frac{A}{\sqrt n})^3(z_u^2-1)\phi(z_u)\nonumber\\
&=&u + [\frac{A}{\sqrt n}+\frac1n(-\frac12z_uA^2+B)\nonumber\\
&\qquad\quad+&\frac1{n^{3/2}}(-\frac16A^3+\frac16A^3z_u^2-ABz_u+C)]\phi(z_u)\nonumber
\end{eqnarray}

The second term of (\ref{eq10})

\begin{eqnarray}
&&\frac1{\sqrt n}p_1(z_u + \frac{A}{\sqrt n} +\frac{B}{n})\phi(z_u+\frac{A}{\sqrt n}+\frac{B}{n})\label{eq12}\\
&=& \frac1{\sqrt n}(-\frac1{12}\eta_3((z_u + \frac{A}{\sqrt n} +\frac{B}{n})^2+2))[1-z_u(\frac{A}{\sqrt n} +\frac{B}{n})+\frac12(z_u^2-1)(\frac{A}{\sqrt n})^2]\phi(z_u)\nonumber\\
&=& \Bigl[\frac1{\sqrt n}(-\frac1{12}\eta_3z_u^2-\frac16\eta_3)+\frac{z_u^3A\eta_3}{12n}+\frac{(-\frac1{24}z_u^4A^2\eta_3+\frac18z_u^2A^2\eta_3+\frac1{12}z_u^3B\eta_3)}{n^{3/2}}\Bigr]\phi(z_u)\nonumber
\end{eqnarray}

The third term of (\ref{eq10})

\begin{eqnarray}
&&\frac1n p_2(z_u + \frac{A}{\sqrt n})\phi(z_u+\frac{A}{\sqrt n}) \label{eq13}\\
&=& \frac1n (-\frac1{288}\eta_3^2 (z_u + \frac{A}{\sqrt n})^5+(\frac1{72}\eta_3^2+\frac18+\frac5{72}\eta_4-\frac16\eta_2)(z_u + \frac{A}{\sqrt n})^3\nonumber\\
&\quad+&(\frac18+\frac1{24}\eta_3^2-\frac1{24}\eta_4)(z_u + \frac{A}{\sqrt n})[1-z_u(\frac{A}{\sqrt n})]\phi(z_u)\nonumber\\
&=& \Bigl[\frac1n (-\frac1{288}\eta_3^2z_u^5+(\frac18+\frac1{72}\eta_3^2+\frac5{72}\eta_4-\frac16\eta_2)z_u^3+(\frac18+\frac1{24}\eta_3^2-\frac1{24}\eta_4)z_u)\nonumber\\
&\quad+& \frac1{n^{3/2}}(\frac1{288}z_u^6\eta_3^2+(-\frac5{72}\eta_4+\frac16\eta_2-\frac1{32}\eta_3^2-\frac18)z_u^4\nonumber\\
&\qquad+&(\frac14+\frac14\eta_4-\frac12\eta_2)z_u^2+\frac18+\frac1{24}\eta_3^2-\frac1{24}\eta_4)A\Bigr]\phi(z_u)\nonumber
\end{eqnarray}

The last term of (\ref{eq10})

\begin{eqnarray}
&&\frac1{n^{3/2}}p_3(z_u)\phi(z_u)\label{eq14}\\
=&&(-\frac1{10368}\eta_3^3z_u^8+(\frac1{96}\eta_3+\frac{19}{10368}\eta_3^3-\frac1{72}\eta_3\eta_2+\frac5{864}\eta_4\eta_3)z_u^6\nonumber\\
&+&(\frac{19}{1728}\eta_3^3-\frac1{30}\eta_5+\frac18\eta_6-\frac1{72}\eta_4\eta_3)z_u^4+(-\frac5{96}\eta_4\eta_3+\frac{35}{864}\eta_3^3+\frac1{32}\eta_3+\frac1{80}\eta_5)z_u^2\nonumber\\
&+& \frac{35}{432}\eta_3^3-\frac5{48}\eta_4\eta_3+\frac1{40}\eta_5+\frac1{16}\eta_3)\phi(z_u)\nonumber
\end{eqnarray}

We take all the terms of the order $1/\sqrt n$ together to find $A$.

\begin{equation} A =\frac{\eta_3}{12}(z_u^2+2).
 \end{equation}

Next, we take all the terms of order $1/n$, plug in the found $A$, to get $B$

\begin{equation}
B= (-\frac18-\frac1{72}\eta_3^2-\frac5{72}\eta_4+\frac16\eta_2)z_u^3+(-\frac1{36}\eta_3^2-\frac18+\frac1{24}\eta_4)z_u
\end{equation}

By plugging $A$ and $B$ in (\ref{eq11}), (\ref{eq12}), (\ref{eq13}), (\ref{eq14}) and taking all the $n^{-3/2}$ terms together we find

\begin{eqnarray}
C &=&(-\frac1{48}\eta_3-\frac1{144}\eta_4\eta_3+\frac1{24}\eta_3\eta_2+\frac1{30}\eta_5-\frac18\eta_6-\frac{19}{1728}\eta_3^3)z_u^4\\
&\quad+&(-\frac5{48}\eta_3+\frac1{12}\eta_3\eta_2-\frac1{80}\eta_5-\frac{67}{1296}\eta_3^3+\frac1{48}\eta_4\eta_3)z_u^2\nonumber\\
&\quad -&\frac1{12}\eta_3-\frac1{40}\eta_5+\frac19\eta_4\eta_3-\frac{113}{1296}\eta_3^3.\nonumber
\end{eqnarray}

\enddocument